\documentstyle[psfig,conf_iap,10pt]{article}
\begin{document}
\heading{%
Galaxy Clusters in the Hubble Volume Simulations
} 
\par\medskip\noindent
\author{%
J\"org M.\ Colberg$^{1,9}$, Simon D.M.\ White$^1$, Thomas J.\ MacFarland$^{2,10}$,
Adrian Jenkins$^3$, Carlos S.\ Frenk$^3$, F.R.\ Pearce$^3$,
August E.\ Evrard$^4$, H.M.P.\ Couchman$^5$, George Efstathiou$^6$,
John A.\ Peacock$^7$, Peter A.\ Thomas$^8$
}
\address{%
Max--Planck--Institut f\"ur Astrophysik, Karl--Schwarzschild--Str.\ 
1, D--85740 Garching, Germany
}
\address{%
Rechenzentrum Garching, Boltzmannstr.\ 2, D--85740 Garching, Germany
}
\address{%
Physics Dept., University of Durham, Durham DH1 3LE, UK
}
\address{%
Physics Dept., University of Michigan, Ann Arbor, MI 48109--1120, USA
}
\address{%
Dept.\ of Astronomy, University of Western Ontario, London, Ontario N6A 3K7, Canada
}
\address{%
Institute of Astronomy, Madingley Road, Cambridge CB3 0HA, UK
}
\address{%
Royal Observatory, Institute for Astronomy, University of Edinburgh, 
Edinburgh EH9 3HJ, UK
}
\address{%
CPES, University of Sussex, Brighton BN1 9QH, UK
}
\address{%
from Oct. 1$^{\rm st}$ 1998 at Cap Gemini GmbH, Kaulbachstr.\ 1, 
D--80539 M\"unchen, Germany
}
\address{%
Now at Enterprise Architecture Group, Global Securities Industry
Group, 1185 Avenue of the Americas, New York 10036, USA
}

\begin{abstract}
We report on analyses of cluster samples obtained from the Hubble
Volume Simulations. These simulations, an $\Omega=1$ model named
$\tau$CDM and a flat low $\Omega$ model with a cosmological
constant ($\Lambda$CDM), comprise the largest computational efforts to date in 
numerical cosmology.  
We investigate the presence of massive galaxy clusters at $z\approx 
0.8$. The $\tau$CDM model fails to form clusters at such a redshift.
However, due to the small number of observed clusters around 
$z\approx 0.8$ and the uncertainties in the determinations of their
masses, this conclusion still is somewhat preliminary. We produce cluster
catalogs at $z=0$ for both
cosmologies and investigate their two--point correlation function $\xi$.
We show that the relationship between the mean density of subsamples
of clusters, expressed via their mean separation $d_{\rm c}$, and
the correlation length $r_0$, defined through $\xi(r_0) = 1$,
is not linear but turns over gently for large $d_{\rm c}$. An analytic
prediction by Mo \& White \cite{Mo} overpredicts $r_0$. The results
from the analysis of the APM cluster data by Croft et al.\ \cite{Croft}
are nicely matched by the $\Lambda$CDM model.
\end{abstract}
\section{The Hubble Volume Simulations}
The Hubble Volume Simulations \cite{Evrard} comprise the
biggest computational effort in numerical cosmology to date.
They follow the evolution of $10^9$ particles in volumes
which contain a significant fraction of the whole observable
Universe and which are several times larger than the largest
of the forthcoming galaxy surveys. Table 1 gives the parameters
of the two models, $\tau$CDM and $\Lambda$CDM. $L$ is the
size of the cubic box in one dimension. Details about the
models can be found elsewhere \cite{Jenkins98a}.

\begin{center}
\begin{tabular}{lcccccc}
\multicolumn{6}{l}{{\bf Table 1.} The Simulation Parameters} \\
\hline
\\
\multicolumn{1}{c}{Model} & \multicolumn{1}{c}{$\Omega$} &
\multicolumn{1}{c}{$\Lambda$} & \multicolumn{1}{c}{$h$} &
\multicolumn{1}{c}{$\Gamma$} & \multicolumn{1}{c}{$\sigma_8$} &
\multicolumn{1}{c}{$L$} \\
\multicolumn{1}{c}{} & \multicolumn{1}{c}{} &
\multicolumn{1}{c}{} & \multicolumn{1}{c}{} &
\multicolumn{1}{c}{} & \multicolumn{1}{c}{} &
\multicolumn{1}{c}{$[\mbox{Gpc}/h]$} \\
\hline
\\
$\Lambda$CDM & 0.3 & 0.7 & 0.7 & 0.21 & 0.9 & 3\\
$\tau$CDM & 1.0 & 0.0 & 0.5 & 0.21 & 0.6 & 2\\
\\
\hline
\end{tabular}
\end{center}

\section{Galaxy Clusters in the Hubble Volume Simulations}

\subsection{Galaxy Clusters at $z\approx 0.8$}

These days, the number of observations of clusters at $z\approx 0.8$ is increasing
at an enormous rate (see e.g.\ \cite{Clowe,Donahue,Luppino}). We have
looked for mass concentrations in the simulation outputs at $z=0.78$
which are in the ranges given in \cite{Clowe,Donahue,Luppino}, that is
$(2.5 - 5.5)\times\,10^{14}\,M_{\odot}/h$. Although
measurements of masses of clusters at these high redshifts still
contain significant uncertainties, nevertheless it can be concluded
that the $\tau$CDM model fails to predict objects of such masses.
The number densities of clusters with a mass of around
$1.5\times\,10^{14}\,M_{\odot}/h$ are more than two orders of
magnitudes below the number densities of the three observed clusters,
and we haven't found a single object with $m>2.0\times\,10^{14}\,M_{\odot}/h$.
The $\Lambda$CDM model does not suffer from these problems. However,
more observational data is clearly required to investigate this
model which is strongly favoured by measurements of distant supernovae
(see the contributions of Perlmutter and Leibundgut in these
proceedings) in more detail (see also \cite{Evrard} for a discussion
of rare events). 

\subsection{The Two--Point Correlation Function of Galaxy Clusters}

The two--point correlation function of galaxy clusters has been
controversial for decades now. Since the early seventies,
it has been known that rich galaxy clusters are
more strongly clustered than galaxies. However, the amplitude  of 
the two--point correlation function and its dependence on cluster 
richness have been the subject of controversy.

The correlation function depends on cluster richness. Richer
clusters are rarer, hence their mean space density, $n_{\rm c}$, is
smaller. Usually this is expressed using the mean intercluster
separation $d_{\rm c}=n_{\rm c}^{-1/3}$. Bahcall (e.g.\ 
\cite{Bahcall92a,Bahcall92b}) has argued
that the correlation length, $r_0$, defined via $\xi(r_0)=1$,
scales linearly with $d_{\rm c}$,
\begin{equation}
r_0 = 0.4\,d_{\rm c}\,.
\label{eq:linear}
\end{equation} 
This {\it ansatz} is based on self--similar scaling, that is 
on the assumption that the correlation function is a power law with 
a fixed slope, $\xi(r) \propto r^{-1.8}$. Early measurements of $\xi(r)$
for the Abell cluster sample \cite{Abell58,Abell89}
seemed to confirm this scaling (e.g.\ \cite{Bahcall83,Peacock92}).
However, Peacock \& West \cite{Peacock92} and Efstathiou et al.\
\cite{Efstathiou92} found strong anisotropy signals in the catalogs
which enhanced the clustering amplitude.

Given a set of points, the two--point correlation function
gives the excess probability of finding a pair of points
separated by a distance $r$ compared with a random Poisson process.
The code we use to compute the two--point correlation function
counts the numbers of such pairs as a function of separation
and from that computes $\xi(r)$. 

Starting from a Press--Schechter 
type {\it ansatz}, Mo \& White (1996)
develop an analytical theory to describe the spatial clustering of
haloes. In particular, they find that the two--point correlation
function of Dark Matter haloes of Lagrangian radius $R$ is related
to that of the mass, $\xi_{\rm DM}$, by
\begin{equation}
\xi(r) = b^2(R)\,\xi_{\rm DM}(r)\,,
\label{eq:mo}
\end{equation}
where
\begin{equation} 
b(R) = 1 + \frac{\delta_{\rm c}}{\sigma^2(R)} - \frac{1}{\delta_{\rm c}}\,.
\label{eq:mob}
\end{equation}
Here, $\delta_{\rm c}=1.69$ is the interpolated linear overdensity at
collapse of a spherical perturbation and $\sigma(R)$ is the {\it
rms} mass fluctuation on the scale of the halo (using a Top--Hat filter).

\begin{figure}
\centerline{\vbox{
\psfig{figure=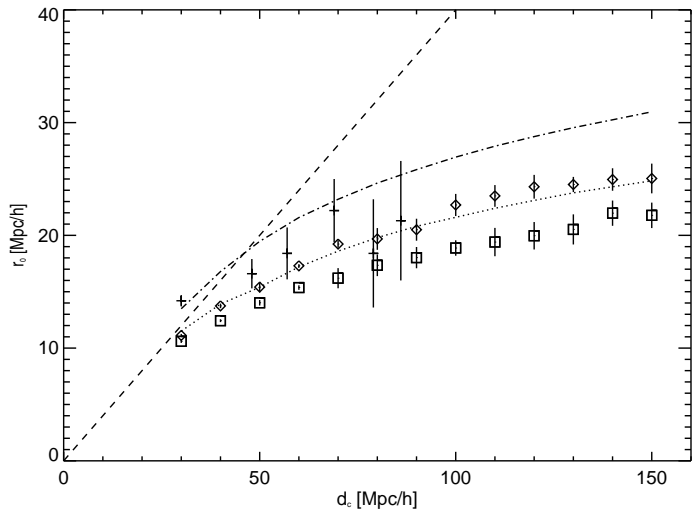,height=6.cm}
}}
\vspace{0.3cm}
\caption[]{The correlation length as a function of the mean intercluster
separation $d_c$ for the $\tau$CDM (boxes with errorbars) and
$\Lambda$CDM (diamonds with errorbars) simulation. The prediction of 
the Mo \& White model are given as dotted and dot--dashed lines for 
the respective simulations. The dashed line is the linear relation 
proposed by Bahcall. Also given are the APM data points taken from 
Croft et al.\ \cite{Croft} (crosses).}
\end{figure}

In order to compute the correlation length, we fit the correlation
function at the vicinity of $\xi(r) = 1$ with a power law of the
following form
\begin{equation}
\xi(r) = \left( \frac{r}{r_0} \right)^{\gamma}\,,
\end{equation}
where $\gamma$ and $r_0$ are free parameters. Figure 1
gives the results of the fits for the correlation lengths. The 
boxes and diamonds are for the $\tau$CDM and $\Lambda$CDM simulation,
respectively. The dotted and dot--dashed line are the results for
the Mo \& White {\it ansatz} using eq.\ (\ref{eq:mo}). Also given
are the results of Croft et al.'s analysis of the APM clusters and
the linear scaling, eq.\ (\ref{eq:linear}), proposed by Bahcall.
The linear scaling fails to reproduce the relation between the
cluster sample density and the correlation length completely. 
Interestingly, the analytical prediction by Mo \& White 
lies above the simulation results. The APM clusters analyzed in 
Croft et al.\ follow the trend of the simulated clusters 
but have slightly larger amplitudes. Clearly, of the two cosmological 
models discussed here, $\Lambda$CDM is more consistent with 
the APM data.

\acknowledgements{JMC thanks Neta Bahcall and Douglas Clowe for 
interesting discussions on the above topics during the conference.}

\vfill
\end{document}